\documentstyle[prl,aps]{revtex}
\begin{document}

\vskip 4mm

\centerline{\large \bf Clusters of interstitial carbon atoms}
\centerline{\large \bf near the graphite surface as a possible origin}
\centerline{\large \bf of dome-like features observed by STM}

\vskip 2mm

\centerline{V.F. Elesin and L.A. Openov}

\vskip 2mm

\centerline{\it Moscow State Engineering Physics Institute
(Technical University),}
\centerline{\it Moscow, 115409, Russia}

\vskip 8mm

\centerline{\bf Abstract}
\begin{quotation}

Formation of clusters of interstitial carbon atoms between the surface and
second atomic layers of graphite is demonstrated by means of molecular
dynamics simulations. It is shown that interstitial clusters result in the
dome-like surface features that may be associated with some of the hillocks
observed by STM on the irradiated graphite surface.

\end{quotation}

\vskip 16mm

{\bf 1. Introduction}

\vskip 2mm

Ion irradiation of graphite surface results in formation of defects that are
seen by scanning tunneling microscopy (STM) \cite{Porte1,Coratger,Porte2,%
Kang,Marton1,Reimann,Bolse,Hahn,Matsukawa}. A single-ion impact creates a
hillock whose nature still remains controversial since STM probes the
electronic states of the surface rather than the actual surface topography.
So, it is difficult to deduce the atomic structure in the vicinity of defects
based on STM data alone.

A number of suggestions have been made concerning the interpretation of
hillocks seen by STM on graphite surfaces irradiated by low-energy
($\sim 100$ eV) ions, including, e.g., the incident ions trapped between the
surface and second layer of atoms in graphite \cite{Kang,Marton1}, the
vacancies on the graphite surface \cite{Hahn,Lee,Nordlund}, self-atom
implantation \cite{Nordlund}, etc. It seems, however, that in many cases the
formation of hillocks involves carbon atoms only.

Recent molecular dynamics simulations \cite{Nordlund} indicate that carbon
interlayer interstitials below the surface layer account for $\approx$ 50
$\%$ of all defects produced by incident and secondary recoils with the
energy $E = 300$ eV $\div$ 3 keV, while $\approx 100$ $\%$ of defects
produced by secondary recoils with the energy $E = 100$ eV $\div$ 300 eV are
single carbon interstitials. It was noticed in \cite{Nordlund} that the
single interstitials may migrate to form clusters between the layers,
which may be the source of some of the hillocks. The authors of
\cite{Nordlund} arrived at a conclusion that, at most, half of the surface
defects may be formed by interstitial carbon clusters.

As to high-energy ($\sim 100$ keV) ion implantation, the hillock shape of the
defects was attributed by several authors to the formation of single carbon
interstitials and interstitial carbon clusters between the graphite layers,
each hillock being originated from a single ion impact \cite{Reimann,Bolse}.
Two types of hillocks were found in \cite{Reimann,Bolse}, one of which
showing the undisturbed (i.e. definitely without vacancies) lateral atomic
arrangement on the (0001) graphite surface, while the other one exhibited a
distorted atomic structure (but also presumably without vacancies). So, it is
likely that the hillocks formed under both low- and high-energy ion
irradiation of graphite are mainly due to interlayer carbon interstitials
and/or their clusters. If this explanation is true, a unique opportunity
appears to study the processes of cluster formation and a related phenomenon
of radiation-induced swelling of graphite by means of STM. Besides, there is
a need for a theory of formation, structure, and arrangement of such
clusters. The theory should give an answer to several key questions: 1) what
is the reason for attraction of interlayer interstitials; 2) what are the
physical conditions of cluster formation; 3) why the recombination of
interstitials with vacancies does not inhibit the formation of clusters.

The purpose of this paper is to study numerically the formation of clusters
of interstitial carbon atoms in the interlayer region between the surface and
second graphite layers. Recently we have
shown by means of molecular dynamics simulations that interlayer carbon
interstitials do form clusters in the bulk of graphite \cite{Elesin1}. The
physical reason for this effect \cite{Elesin2} is the deformation interaction
of interstitials mediated by the nearby graphite sheets. An interstitial
stretches the lattice in its vicinity, and the stretched domain appears to be
attractive for other interstitials located within the same interlayer region,
see Fig.1. In other words, the formation of interstitial clusters minimizes
the deformation energy relatively to the case of spatially separated
interstitials due to minimization of the overall buckling of adjacent
graphite sheets. It is important that mobility of interstitial clusters is
much lower than that of single interstitials, and hence the recombination
of interstitials with vacancies is strongly suppressed \cite{Elesin3}.

On general grounds, one can expect the same mechanism of clusters formation
to be realized near the graphite surface. Indeed, the formation of clusters
of noble-gas atoms inserted between the topmost and second atomic layers of
graphite has been numerically demonstrated by Marton et al.
\cite{Marton1,Marton2}. We are not aware of similar studies on the clusters
of carbon atoms.

The paper is organized as follows. First we present the general theory of
diffusion instability of the uniform distribution of defects in solids.
Next we numerically simulate the formation of interstitial carbon clusters
underneath the surface layer of graphite. Our calculations show that such
clusters can indeed be formed. Their specific shapes are dictated by the
initial distribution of interstitials in the interlayer region and by the
strong covalent interaction between the interstitials. The interstitial
clusters give rise to dome-like features on the graphite surface that are
similar to those observed experimentally by STM.

\vskip 6mm

{\bf 2. Diffusion instability of a system of defects in solids}

\vskip 2mm

In this Section we sketch the main features of the theory of diffusion
instability that can lead to formation of clusters of defects in disordered
solids, see \cite{Elesin2,Elesin3} for more details. We shall demonstrate
that the uniform distribution of defects is unstable with respect to
separation in defect-rich and defect-poor regions if the concentration of
defects is sufficiently high and exceeds some critical value that depends on
material parameters and temperature. The mechanism of instability is rather
general: the deformation of the crystal lattice by defects results in
anomalous flow of defects along the gradient of defect concentration
$n({\bf r},t)$, i.e. opposite in direction to the diffusion flow of defects.

In the simplest case of defects of the same type, the kinetic equation for
$n({\bf r},t)$ is
\begin{equation}
\frac{\partial n}{\partial t}=Q-R(n)-div{\bf j} ,
\label{Equation1}
\end{equation}
where $Q$ is the source of defects, the term $R(n)$ accounts for
recombination processes, and ${\bf j}$ is the defect current density. In the
case of isotropic continuum media, the expression for ${\bf j}$ reads
\begin{equation}
{\bf j}=-D\frac{\partial n}{\partial {\bf r}}+
\frac{D\Omega n}{3T}\frac{\partial \sigma_{ii}}{\partial {\bf r}} ,
\label{Equation2}
\end{equation}
where the first term corresponds to the diffusion of defects, while the
second one describes the motion of defects in the field of elastic stresses
$\sigma_{ii}=div{\bf \sigma}$. Here $D$ is the diffusion coefficient and
$\Omega$ is the dilatation volume ($\Omega > 0$ for interstitials and
$\Omega < 0$ for vacancies; $|\Omega| \sim a^3$, where $a$ is the lattice
constant).

In turn, average elastic stresses $\langle\sigma_{ii}\rangle$ caused by
defects satisfy the following equation of the theory of elasticity:
\begin{equation}
\frac{\partial \langle\sigma_{ii}\rangle}{\partial {\bf r}}=
3K^*\Omega\frac{\partial n}{\partial {\bf r}} ,
\label{Equation3}
\end{equation}
where
\begin{equation}
K^*=K\left(1+\frac{4\mu}{3K}\right)^{-1} ,
\label{Equation4}
\end{equation}
$K$ and $\mu$ are elastic moduli. Substituting Eq.(\ref{Equation3}) in
Eq.(\ref{Equation2}), we obtain the following expression for ${\bf j}$:
\begin{equation}
{\bf j}=-D\left(1-\frac{\Omega^2K^*n}{T}\right)
\frac{\partial n}{\partial {\bf r}} .
\label{Equation5}
\end{equation}
It follows from Eq.(\ref{Equation5}) that the current caused by elastic
stresses is directed opposite to the diffusion component of the current. This
result has a simple physical meaning. Let us consider, for example, the
system of defects with $\Omega > 0$. Such defects (e.g., interstitials)
stretch the lattice, i.e. increase the {\it average} lattice constant, see
Eq.(\ref{Equation3}). But since interstitials themselves tend to move to the
stretched regions of the sample, there is an anomalous current along the
gradient of defects concentration. This is what we call the deformation
interaction of defects.

The steady-state spatially-uniform solutions of Eqs. (\ref{Equation1}),
(\ref{Equation5}) appear to be unstable at
\begin{equation}
n>n_c=\frac{T}{\Omega^2K^*} .
\label{Equation6}
\end{equation}
Hence, in the case that the average concentration of defects exceeds the
critical concentration $n_c$, the agglomerations of defects (clusters) are
formed. For $T=300$K and typical values of $\Omega\approx 5\cdot10^{-23}$
cm$^3$ and $K^*\approx 10^{12}$ erg/cm$^3$ we have $n_c\sim 10^{19}$
cm$^{-3}$, this value being much smaller than the atomic concentration.

We stress that the results presented above are obtained within the framework
of a standard defect model in the context of the linear theory of elasticity,
without use of any new assumptions and parameters. Note also that the
diffusion instability is a common feature of a number of other physical
systems, e.g., superconductors and excitonic insulators \cite{Elesin4}.

Our results hold true also for defects with $\Omega < 0$, e.g., vacancies.
Indeed, a vacancy compresses the lattice, i.e. reduce the average lattice
constant, while the compressed domain appears to be attractive for other
vacancies. If both defects with $\Omega > 0$ and $\Omega < 0$ (e.g.,
interstitials and vacancies respectively) are present in the sample, their
effective interaction with each other via the lattice deformation is
repulsive \cite{Elesin2}. Such a repulsion leads to spatial separation of
vacancies and interstitials. Interstitials form the interstitial clusters,
while vacancies form the vacancion clusters. The interstitial clusters are
much less mobile than a single interstitial. As a result, the recombination
is suppressed, and the concentration of defects under the influence of an
external source can increase monotonously for a very long period of time
without saturation \cite{Elesin3}. We stress that our model was formulated
for the case of a {\it finite} defect concentration, and hence does not imply
that, e.g., a {\it single} vacancy and a {\it single} interstitial have a
repulsive interaction.

It should be noted that while in isotropic crystals a characteristic range of
deformation interaction is very short, about one lattice constant (and hence
such an interaction can be viewed as point-like in the continuum limit
\cite{Elesin2,Elesin3}), in anisotropic crystals the range of deformation
interaction can be much longer. In particular, interlayer interstitials in
graphite (both impurities and carbon atoms) are known to deform atomic layers
very strongly \cite{Marton2,Taji}. As a corollary, the deformation
interaction between the defects in graphite should play an important role in
temporal evolution of the system of defects under irradiation and influence
the spatial arrangement of defects over the sample. The formation of
spatially separated interstitial and vacancy clusters and the corresponding
suppression of recombination can explain some peculiar features of graphite
swelling under irradiation \cite{Elesin2,Elesin3}. Note that other mechanisms
(e.g., the specific bonding characteristics of the defects) can also
contribute to the recombination suppression in graphite.

The effect of interstitial carbon atoms formation in the bulk of graphite has
been demonstrated numerically in \cite{Elesin1} making use of empirical
interatomic potentials. Below we study the formation of such clusters beneath
the surface graphite layer taking into account the strong covalent bonding
between carbon atoms constituting the cluster.

\vskip 6mm

{\bf 3. Computational details}

\vskip 2mm

Graphite was modeled by six layers of hexagonal carbon networks, each
layer staggered with respect to its neighbors, see Fig.2. The layers were
composed of 220 atoms each. The atoms at the periphery of each layer were
fixed in order to keep the stability of the crystallite. All other atoms were
allowed to move after the interstitials have been inserted into the
crystallite. Initially ($t=0$) ten interstitial carbon atoms were positioned
randomly between one of the peripheral layers (the "surface layer") and the
penultimate layer.

In the presence of interstitials, the undisturbed graphite layers correspond
to a highly nonequilibrium atomic configuration. The potential energy of the
system rapidly decreases in time, leading to an increase in the kinetic
energy. An additional decrease of the potential energy occurs because of the
formation of covalent bonds between the interstitial carbon atoms upon fusion
of interstitials into clusters. In order to avoid the overheating of the
crystallite, we eliminated the whole kinetic energy of atoms artificially
each 70 MD steps (each period of time $t_o=2\cdot 10^{-14}$ s), i.e. in fact
we removed the heat from the crystallite. A rough estimate of the system
temperature made on the basis of values of the maximum kinetic energy
attained each period $t_o$, gives $T\sim 1000$ K at the very beginning of
simulations and $T\sim 10$ K at the final stage (after the formation of a
single interstitial cluster).

We made use of empirical interatomic potentials proposed by Taji et al.
\cite{Taji} to account for interaction between the carbon atoms located in
graphite layers as well as between the interstitial carbon atoms and the
atoms of graphite layers. Parameters of those potentials were chosen by
fitting the calculated values of interatomic distances and elastic moduli to
their experimental values in defect-free graphite. There are two parts in the
Taji potential. The first part describes the interaction between the atoms
{\it within} a graphite layer and takes the non-central force interaction
into account in the form of the Keating strain energy potential. This part
has an explicit minimum for angles between C-C bonds of 120$^o$ and hence
treats $sp^2$ bonding. The second part describes the interaction
{\it between} graphite layers by the central force potential. This part
includes the attractive van der Waals force and the repulsive force that
prevents graphite layers from merging together. The second part of the
potential was applied by Taji et al. to the interactions not only between
the carbon atoms located on the different layers but also between the
interstitial carbon atom and other lattice atoms. For an interstitial located
between graphite layers, its interaction with graphite layers, as described
by the Taji potential, appears to be {\it repulsive}, i.e. the interstitial
atom does not form covalent bonds with carbon atoms of the nearby graphite
layers. Such a choice of the potential agrees with the experimentally
observed expansion of graphite due to interstitials that gives evidence for
the {\it repulsion} between interstitials and graphite layers, i.e. for the
absence of covalent bonds between the interstitial atom and the graphite
layers. In other words, the interstitial self-energy $E_{is}$ is positive.
Its value $E_{is}=$ 1.8 eV calculated using the Taji potential agrees well
with experimental data and theoretical estimates \cite{Coulson,Thrower}.

Note that the Taji potential was originally proposed \cite{Taji} for the case
that the interatomic distances and bond angles are not so far from their
equilibrium values in graphite. This condition is, however, fulfilled in the
simulations of formation of both a single interstitial \cite{Taji} and
interstitial clusters since the distortion of the surface layer by
interstitials is rather smooth, see below.

Of course, the Taji potential is empirical and its use should be also
justified by more sophisticated calculations. Since first principles
calculations of interstitial characteristics in graphite are not known to us,
we have carried out TBMD (tight binding molecular dynamics \cite{Xu1,Xu2})
simulations of interstitials between the graphite layers. We have
demonstrated that there is indeed a {\it repulsive} force between an
interstitial and the nearby graphite layers, so that new covalent bonds are
not formed and $sp^3$ bonding does not appear. We have also shown that there
is not only qualitative, but also semi-quantitative agreement between TBMD
calculations and simulations based on the Taji potential. In particular, the
calculated values of the volume expansion due to an interstitial, the
displacements of atoms in the nearest to an interstitial graphite layers, and
the value of $E_{is}$ are close to those found using the Taji potential.

In order to account for the covalent interaction between the interstitials
we employed the TBMD method that had been proven to give a good description
of small carbon clusters \cite{Xu1,Xu2,Openov}.

\vskip 6mm

{\bf 4. Results and discussion}

Fig.3 shows the dynamics of 10-interstitial cluster formation in the
interlayer region nearest to the graphite surface. Initially ($t=0$) the
distance between any two interstitials greatly exceeds the characteristic
range of covalent interaction ($\approx$ 2 {\AA}). At the first stage the
motion of interstitials is governed entirely by the attractive deformation
interaction \cite{Elesin1}. At $t=100$, a small 2-interstitial cluster is
formed. The atoms of the cluster are tightly bound together by covalent
bonding, the covalent component of the binding energy being
$E_b=$ 2.85 eV/atom, in agreement with the experimental value for the dimer
C$_2$ \cite{Huber}. At $t=200$, one more 2-interstitial cluster is formed. In
fact, at $t = 200 \div 500$ two processes take place, the formation of new
diatomic clusters and adsorption of single interstitials by those clusters.
The main reason for the motion of single interstitials towards the
interstitial clusters is the deformation-induced attraction \cite{Elesin1},
as in the case of 2-interstitial clusters formation. At $t=500$, there are
two 3-interstitial clusters and one 4-interstitial cluster in the interlayer
region. Those clusters have the linear chain structure governed by the
covalent interaction between the interstitials within the clusters
\cite{Xu2}. The covalent component of the binding energy per interstitial
is $E_b=$ 4.73 eV/atom and $E_b=$ 4.75 eV/atom for 3- and 4-interstitial
clusters respectively, in accordance with theoretical results of other
authors and experimental values (see, e.g., \cite{Menon,Drowart}).

At the next stage of the evolution ($t = 500 \div 5000$), three interstitial
clusters slowly move towards each other. At $t = 5000$, two clusters merge
into a 6-interstitial cluster. Note that characteristic times of 3- and
4-interstitial clusters formation are an order of magnitude shorter than the
time it takes for those clusters to merge together. In other words, the
mobility of an interstitial cluster decreases rapidly as the number of
interstitials in the cluster increases.

Finally, the 10-interstitial cluster is formed, see Fig.3h. The shape of this
cluster is typical for low-dimensional carbon structures that are
characterized by the bond angles 180$^o$ (carbyne) and 120$^o$ (graphite
layers) and by coordination numbers $Z=2$ and $Z=3$. The covalent component
of the binding energy of interstitials within the cluster is
$E_b=$ 5.50 eV/atom. Note that the most stable geometry of an {\it isolated}
10-atom carbon cluster is a ring structure \cite{Xu2,Raghavachari,Tomanek},
its binding energy calculated using the same TBMD method equals to
$E_b=$ 5.87 eV/atom. Nevertheless, we have verified that the shape and the
binding energy of the 10-interstitial cluster shown in Fig.3h remained nearly
unchanged upon its removal from the crystallite and subsequent relaxation.
Hence this cluster, taken alone (outside the graphite) is metastable and
corresponds to a local energy minimum.

The deformation of the surface layer by the 10-interstitial cluster is shown
in Fig.4. One can see the dome-like feature on the graphite surface. Its
height is about 1.5 {\AA} and lateral dimensions are about $10 \div 12$ {\AA}.
We suggest that some of the hillocks seen by STM on irradiated graphite
surfaces \cite{Porte1,Coratger,Porte2,Kang,Marton1,Reimann,Bolse,Hahn,%
Matsukawa} may be due to relatively large (composed of $\sim$ 10 carbon
atoms) interstitial clusters formed in the interlayer region between the
surface and penultimate graphite layers at the sacrifice of deformation
interaction of interstitials. The case considered by us (10 interstitials are
created simultaneously in the surface area $\sim$ 10 {\AA}$^2$) obviously
corresponds to rather high energy of incident ions. Note, however, that
interstitials' dynamics shown in Fig.3 may be viewed as the {\it final stage}
of cluster formation from well isolated interstitials initially created far
apart from each other by low-energy ions and migrated to the region confined
by the crystallite due to deformation attraction and/or thermally activated
diffusion.

We have made molecular dynamics simulations of interstitial clusters
formation for several other initial distributions of interstitials over the
interlayer region nearest to the graphite surface. The results are shown in
Figs.5-7. They are similar to those presented above. One can see that in
either case studied all interstitials, initially located far apart from each
other, merge into a single 10-interstitial cluster. The shape of the cluster
is not unique, it obviously depends on the initial distribution of
interstitials and hence on the specific sequence of small clusters formation
and subsequent fusion into a single large cluster. However, the angles
between the C-C bonds in those clusters are equal or close to either 180$^o$
or 120$^o$, while the coordination numbers of interstitials within the
clusters are $Z=2$ or $Z=3$ (with obvious exception of boundary interstitials
for which $Z=1$), just as in the cluster shown in Fig.3h. Thus the shape of
the interstitial cluster is strongly influenced by the covalent interactions
between the interstitials within the cluster.

The covalent components of the binding energy of interstitials with each
other equal to $E_b=$ 5.34 eV/atom, 5.32 eV/atom, and 5.68 eV/atom for the
clusters shown in Figs. 5b, 6b, and 7b respectively. Those values of $E_b$
are lower than the binding energy $E_b=$ 5.87 eV/atom of a stable 10-atom
ring. Nevertheless we have found that the clusters preserve their shapes upon
their removal from the crystallite, the values of $E_b$ being changed
insignificantly. Thus all final configurations of interstitials are
metastable.

Distortions of the surface layer by 10-interstitial clusters are shown in
Figs. 5c, 6c, and 7c. It is clearly seen that the shape of the dome-like
feature on the graphite surface reflects the particular shape of interstitial
cluster. The height of the hillock varies slightly from one interstitial
configuration to another, its typical value being 1.5 {\AA}. The lateral
dimensions of the hillock are more sensitive to the shape of the cluster,
usually being in the range 10 $\div$ 15 {\AA}.

Finally, we have also simulated the deformation of the graphite surface by a
single carbon interstitial and by interstitial clusters composed of various
numbers of interstitial carbon atoms. We have found that the calculated
height of the surface hillock increases monotonously with the number of
interstitials in the cluster, starting from about 0.8 {\AA} for a single
interstitial. The same holds true for characteristic lateral dimensions of
the hillock. We believe that such a variation in size of hillocks may be at
least a part of the reason for experimentally observed variety of features on
irradiated graphite surfaces. For example, Coratger et al. \cite{Coratger}
have found that irradiation of HOPG graphite with 20 keV $^{12}$C$^{+}$ ions
resulted in protrusions presented an {\it elongated} form. These finding can
be explained by the formation of relatively small clusters C$_N$ beneath the
graphite surface since for $N<10$ a linear chain structure of carbon clusters
ia believed either to be the most stable geometry or to lie very close in
energy to a ring structure \cite{Xu2,Openov,Menon,Tomanek}.

\vskip 6mm

{\bf 5. Summary and conclusions}

\vskip 2mm

Making use of molecular dynamics simulations, we have demonstrated that
interstitial carbon atoms formed in the interlayer region between the surface
and second graphite layers under ion irradiation attract each other. The
physical origin for such an attraction is the strong deformation of graphite
layers by interstitials that makes it energetically favorable for
interstitials to come closer to each other (and thus to minimize the overall
buckling of graphite layers) than to stay far apart.

Attractive deformation interaction of interstitials results in formation of
interstitial clusters from initially isolated single interstitials. The
specific shape of the cluster depends on the initial distribution of
interstitials over the interlayer region and is dictated by the strong
covalent interaction between the interstitials. Coordination numbers for the
majority of interstitials in the cluster are $Z=2$ and $Z=3$, while all
angles between the C-C bonds in the cluster are close to 180$^o$ or
120$^o$, these values being typical for low-dimensional carbon structures.
Various arrangements of interstitials in the clusters correspond to different
metastable states of the cluster at a fixed number of carbon atoms in it.

Deformation of the surface layer by interstitial clusters results in the
dome-like features on the graphite surface. The typical height and lateral
dimensions of the hillocks produced by 10-interstitial clusters are 1.5 {\AA}
and 10 $\div$ 15 {\AA} respectively. Such features have been observed
repeatedly by STM on the graphite surfaces irradiated with noble gas ions.
The results presented in this paper provide an explanation to experimental
observations. In order to make a closer comparison between the theory and
experiment, it would be interesting to calculate STM images from the hillocks
formed above interstitial clusters of different shapes in graphite.

\vskip 6mm

{\bf Acknowledgments}

\vskip 2mm

The work was supported by the International Science and Technology Center
(Project 467) and by the Russian State Program "Integration".

\newpage
\centerline{\bf FIGURE CAPTIONS}
\vskip 2mm

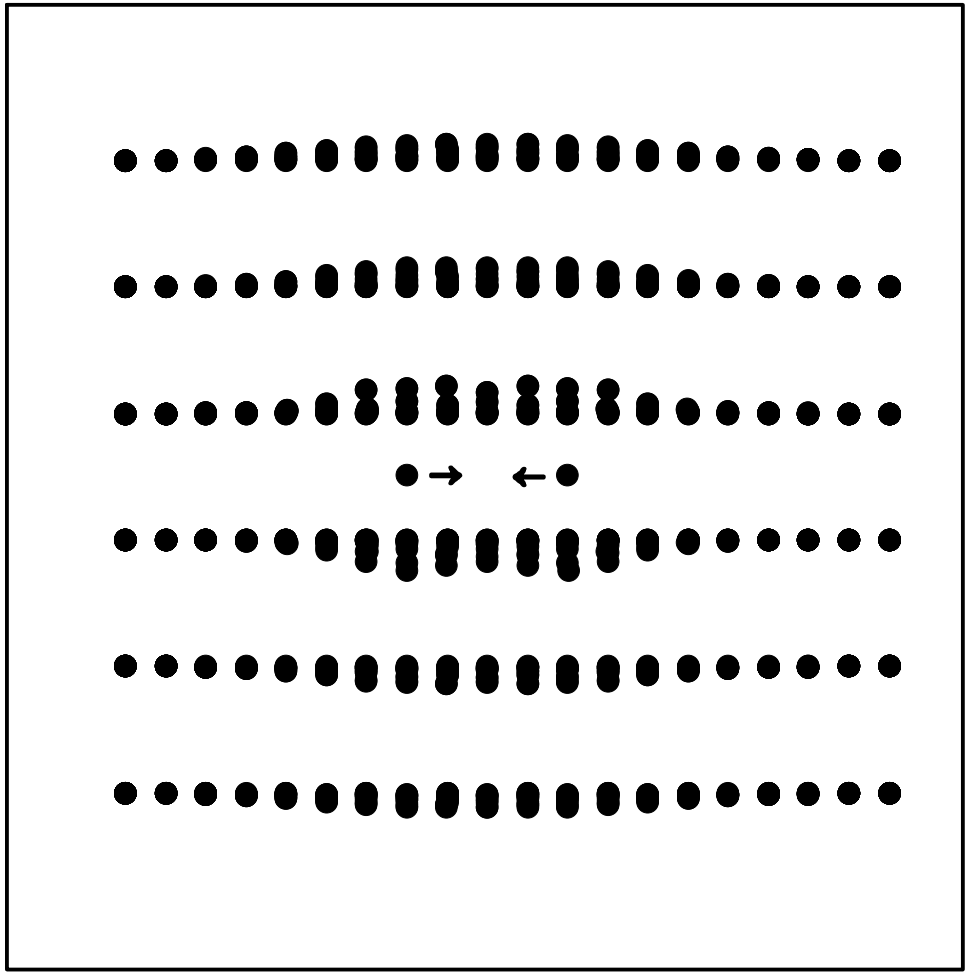
Fig.1. Distortion of graphite lattice by two interstitials located between
the nearest graphite layers at the in-plane distance $r=$ 4.9 {\AA} from each
other.

Fig.2. Side view (a) and top view (b) of the crystallite composed of 1320
carbon atoms (six layers 220 atoms each). Closed and open circles show the
atoms of odd and even (from the top to the bottom) layers respectively.

Fig.3. Dynamics of 10-interstitial cluster formation in the interlayer region
nearest to the graphite surface. Top view. Atoms of graphite layers are not
shown. Time $t$ is measured in units of $t_o=2\cdot 10^{-14}$ s. $t=0$ (a),
$t=100$ (b), $t=200$ (c), $t=300$ (d), $t=500$ (e), $t=2000$ (f),
$t=5000$ (g), $t=20000$ (h).

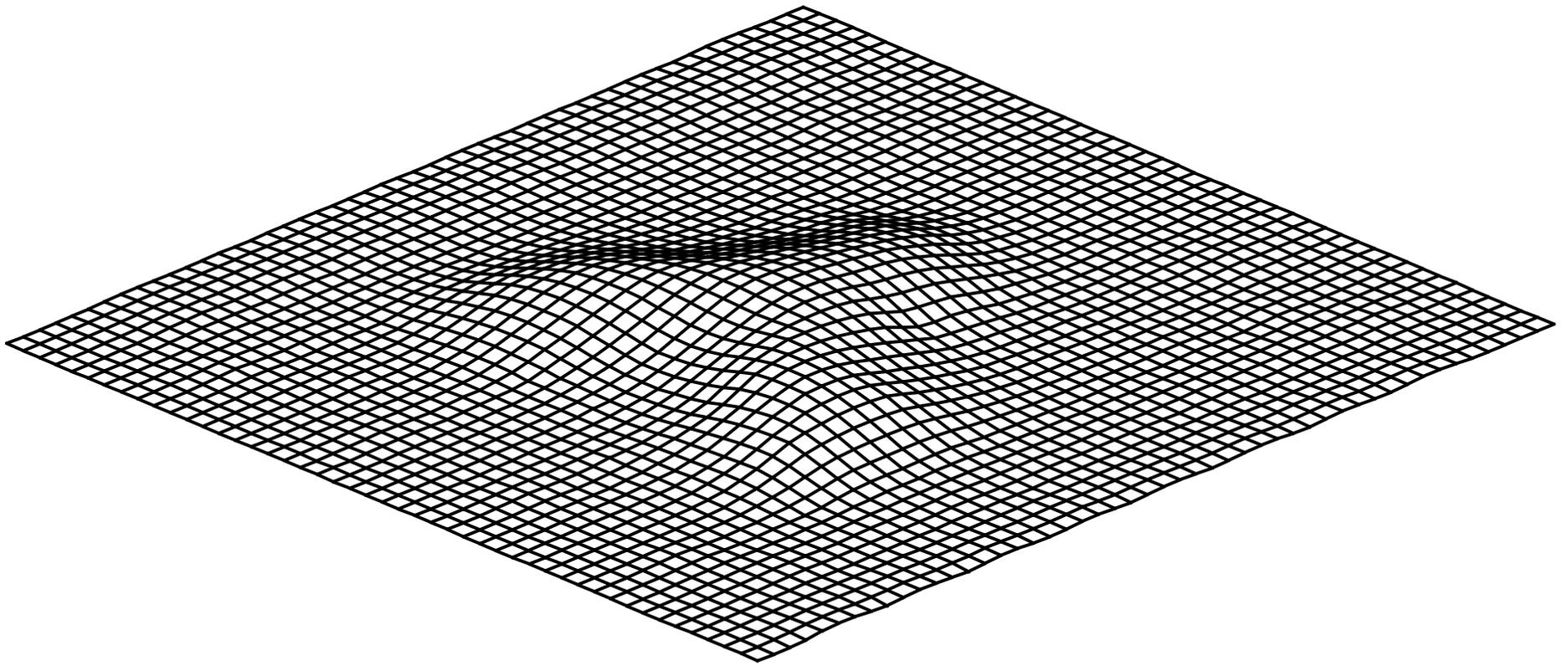
Fig.4. Deformation of the surface layer by the 10-interstitial cluster.

Fig.5. Initial (a) and final (b) arrangement of 10 carbon interstitials in
the interlayer region between the surface and penultimate layers of graphite.
Top view. Atoms of graphite layers are not shown. Deformation of the surface
layer by the 10-interstitial cluster (c).

Fig.6. Same as in Fig.4 for other initial distribution of 10 interstitials
over the interlayer region.

Fig.7. Same as in Fig.4 for other initial distribution of 10 interstitials
over the interlayer region.


\vskip 6mm

\begin{references}

\bibitem{Porte1} L. Porte, M. Phaner, C.H. de Villeneuve, N. Moncoffre and
J.Tousset, Nucl. Instrum. Methods B 44 (1989) 116.

\bibitem{Coratger} R. Coratger, A. Claverie, F. Ajustron and J. Beauvillain,
Surf. Sci. 227 (1990) 7.

\bibitem{Porte2} L. Porte, C.H. de Villeneuve and M. Phaner,
J. Vac. Sci. Technol. B 9 (1991) 1064.

\bibitem{Kang} H. Kang, K.H. Park, C. Kim, B.S. Shim, S. Kim and D.W. Moon,
Nucl. Instrum. Methods B 67 (1992) 312.

\bibitem{Marton1} D. Marton, H. Bu, K.J. Boyd, S.S. Todorov, A.H. Al-Bayati
and J.W. Rabalais, Surf. Sci. 326 (1995) L489.

\bibitem{Reimann} K.P. Reimann, W. Bolse, U. Geyer and K.P.Lieb,
Europhys. Lett. 30 (1995) 463.

\bibitem{Bolse} W. Bolse, K.P. Reimann, U. Geyer and K.P. Lieb,
Nucl. Instrum. Methods B 118 (1996) 488.

\bibitem{Hahn} J.R. Hahn, H. Kang, S. Song and I.C. Jeon,
Phys. Rev. B 53 (1996) R1725.

\bibitem{Matsukawa} T. Matsukawa, S. Suzuki, T. Fukai, T. Tanaka and
I. Ohdomari, Appl. Surf. Sci. 107 (1996) 227.

\bibitem{Lee} K.H. Lee, H.M. Lee, H.M. Eun, W.R. Lee, S. Kim and D. Kim,
Surf. Sci. 321 (1994) 267.

\bibitem{Nordlund} K. Nordlund, J. Keinonen and T. Mattila,
Phys. Rev. Lett. 77 (1996) 699.

\bibitem{Elesin1} V.F. Elesin and L.A. Openov,
Phys. Low-Dim. Struct. 7/8 (1998) 195.

\bibitem{Elesin2} V.F. Elesin, Dokl. Akad. Nauk SSSR, 298 (1988) 1377
[Sov. Phys. Dokl., 33 (1988) 138].

\bibitem{Elesin3} V.F. Elesin, Pis'ma Zh. Eksp. Teor. Fiz., 59 (1994) 451
[JETP Lett., 59 (1994) 472].

\bibitem{Marton2} D. Marton, K.J. Boyd, T. Lytle and J.W.Rabalais,
Phys. Rev. B 48 (1993) 6757.

\bibitem{Elesin4} V.F. Elesin and Yu.V. Kopaev, Usp. Fiz. Nauk 133 (February
1981) 259 [Sov. Phys. Usp. 24 (February 1981) 116].

\bibitem{Taji} Y. Taji, T. Yokota and T. Iwata,
J. Phys. Soc. Jap. 55 (1986) 2676.

\bibitem{Coulson} C.A. Coulson, S. Senent, M.A. Herraez, M. Leal and
E. Santos, Carbon 3 (1966) 445.

\bibitem{Thrower} P.A. Thrower and R.M. Mayer, Phys. Stat. Sol. (a)
47 (1978) 11.

\bibitem{Xu1} C.H. Xu, C.Z. Wang, C.T. Chan and K.M. Ho,
J. Phys.: Condens. Matter 4 (1992) 6047.

\bibitem{Xu2} C.H. Xu, C.Z. Wang, C.T. Chan and K.M. Ho,
Phys. Rev. B 47 (1993) 9878.

\bibitem{Openov} L.A. Openov and V.F. Elesin, Pis'ma v ZhETF 68 (1998) 695
[JETP Lett. 68 (1998) 726].

\bibitem{Huber} K.P. Huber and G. Herzberg, {\it Constants of Diatomic
Molecules}, Van Nostrand Reinhold, New York, 1979.

\bibitem{Menon} M. Menon, K.R. Subbaswamy and M. Sawtarie,
Phys. Rev. B 48 (1993) 8398.

\bibitem{Drowart} J. Drowart, R.P. Burns, G. De Maria and M.G. Inghram,
J. Chem. Phys. 74 (1959) 1131.

\bibitem{Raghavachari} K. Raghavachari and J.S. Binkley,
J. Chem. Phys. 87 (1987) 2191.

\bibitem{Tomanek} D. Tom\'anek and M.A. Schluter,
Phys. Rev. Lett. 67 (1991) 2331.

\end{references}
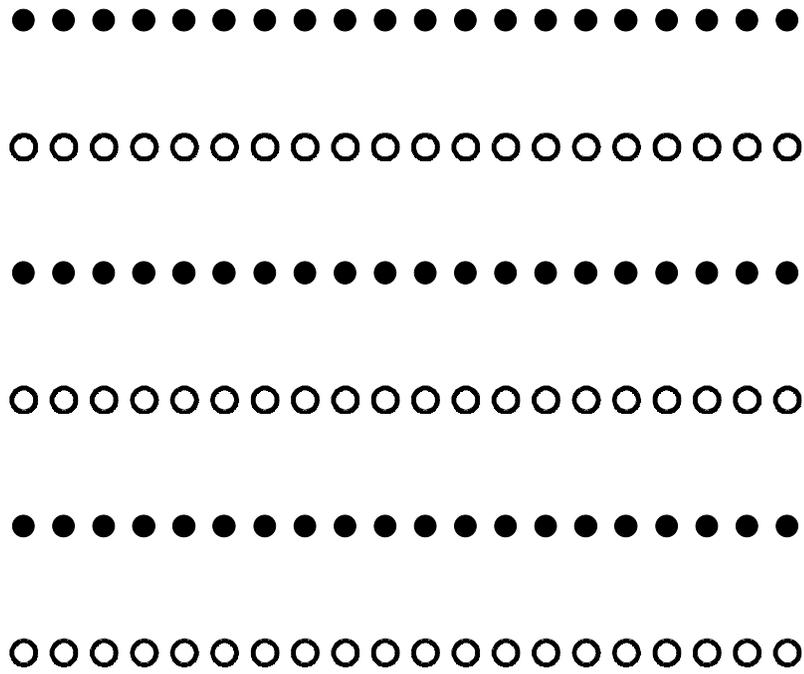
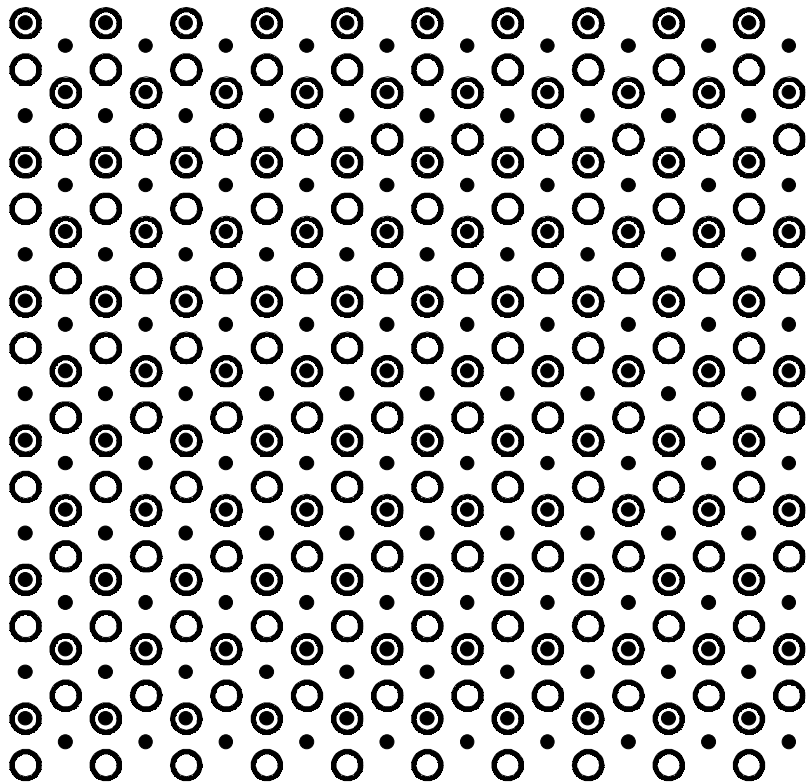
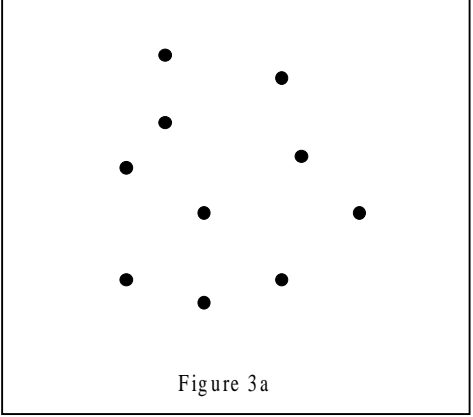
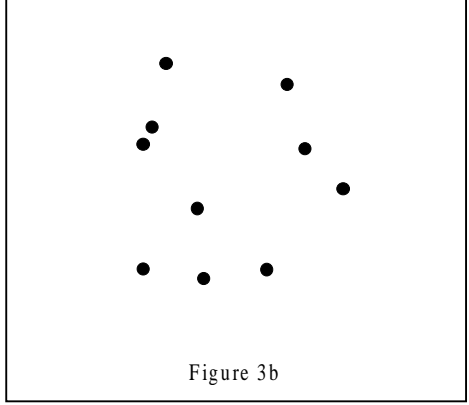
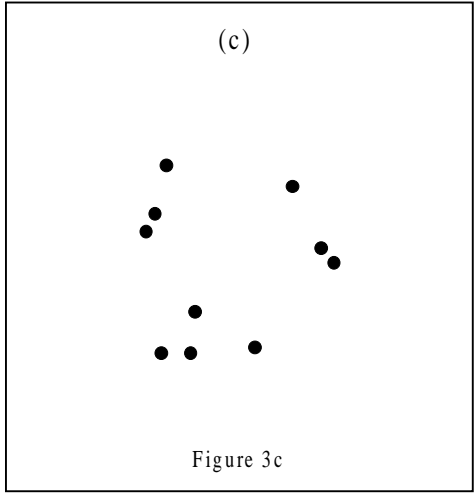
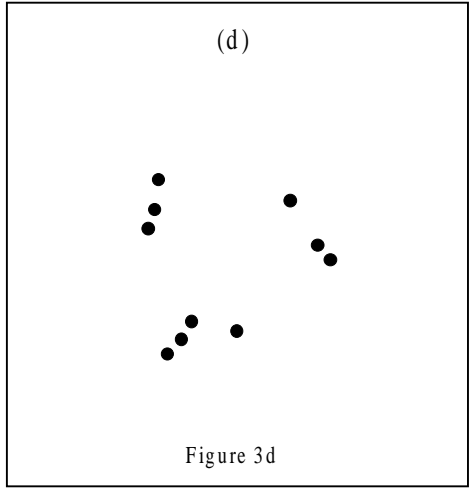
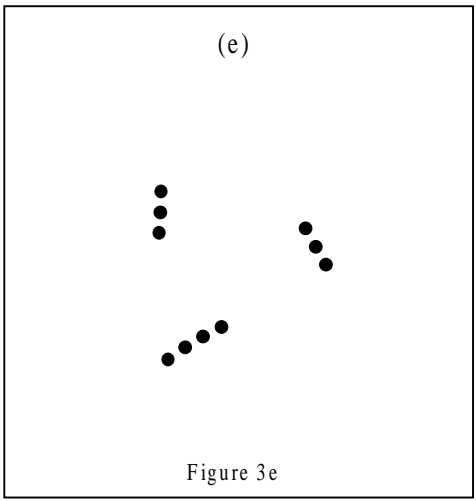
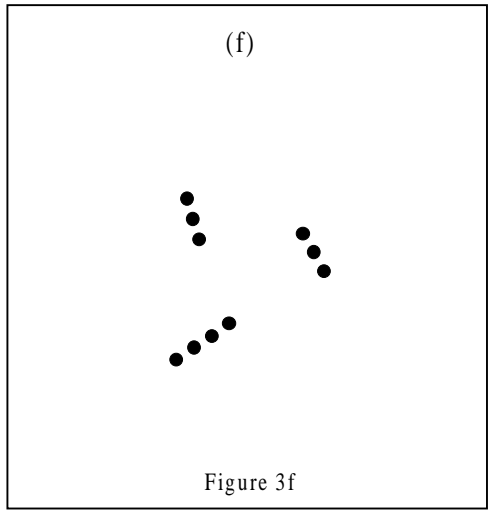
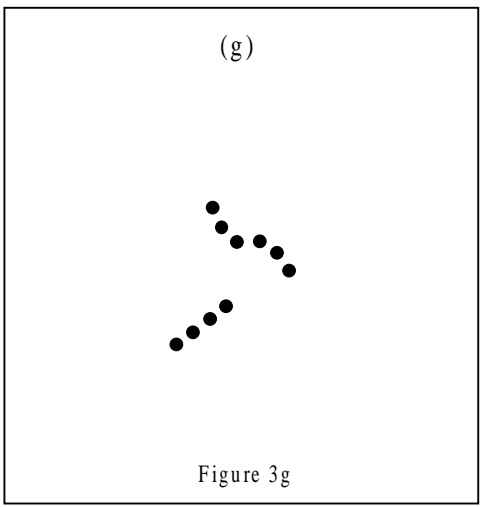
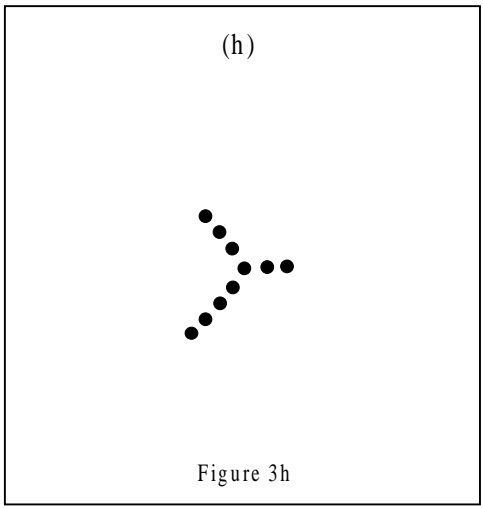
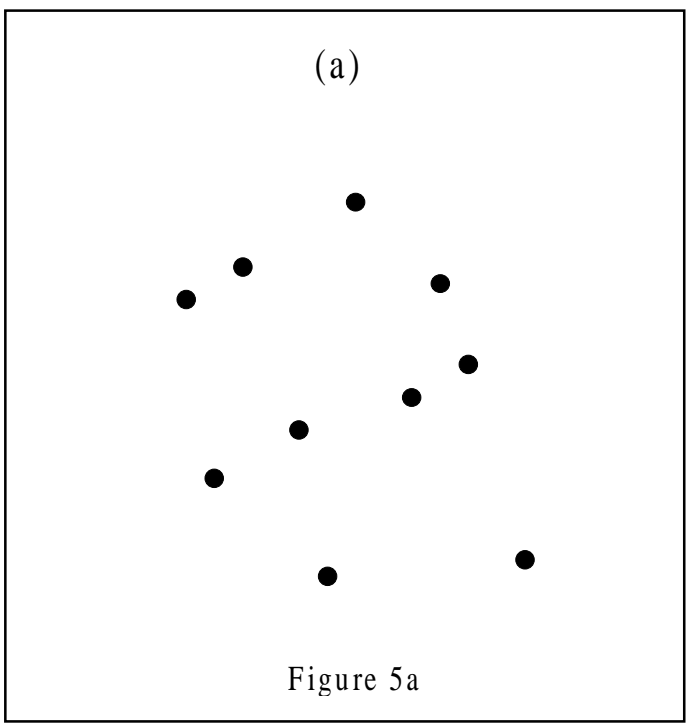
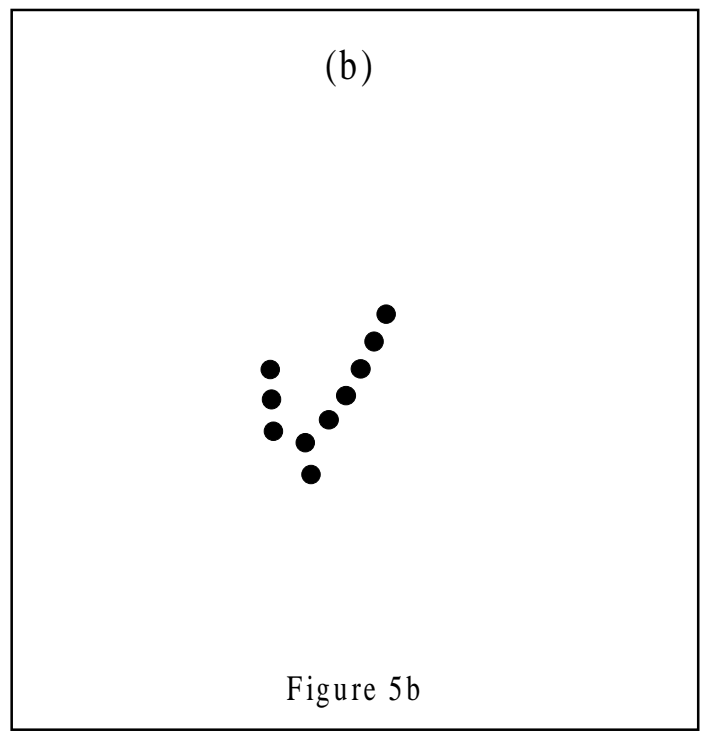
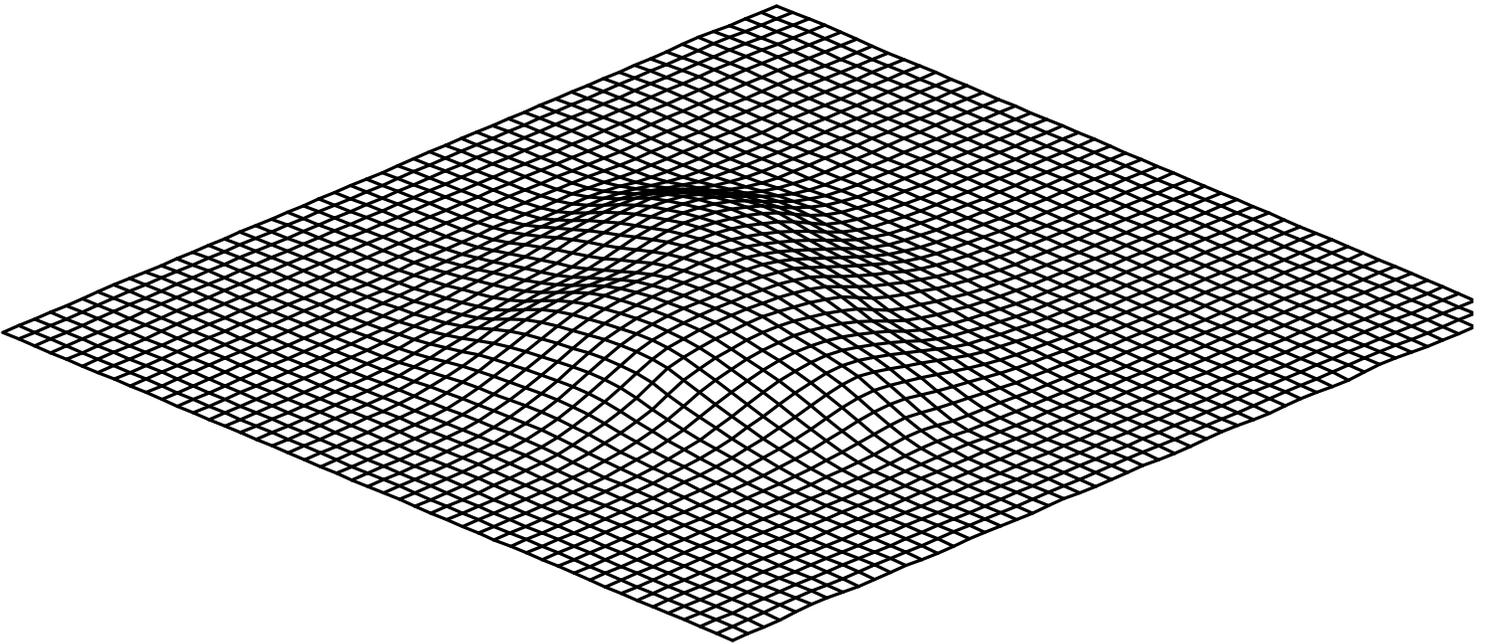
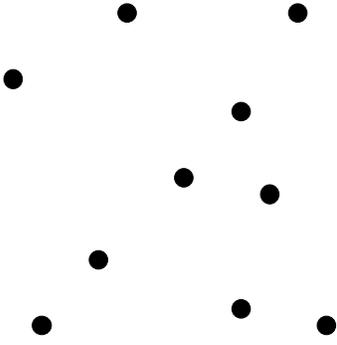
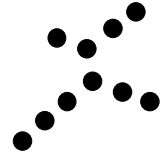
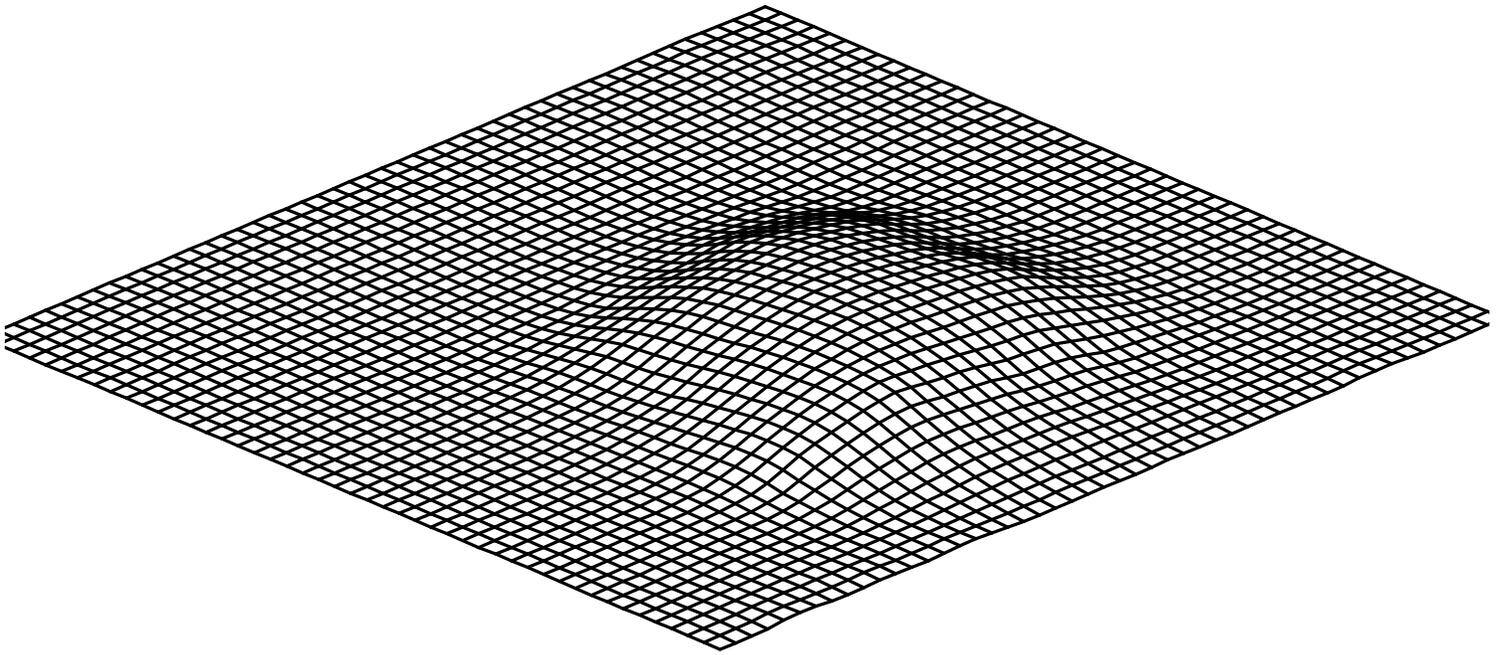
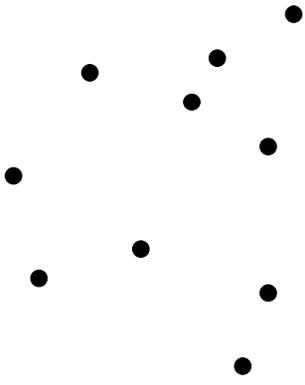
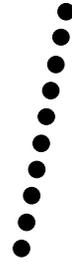
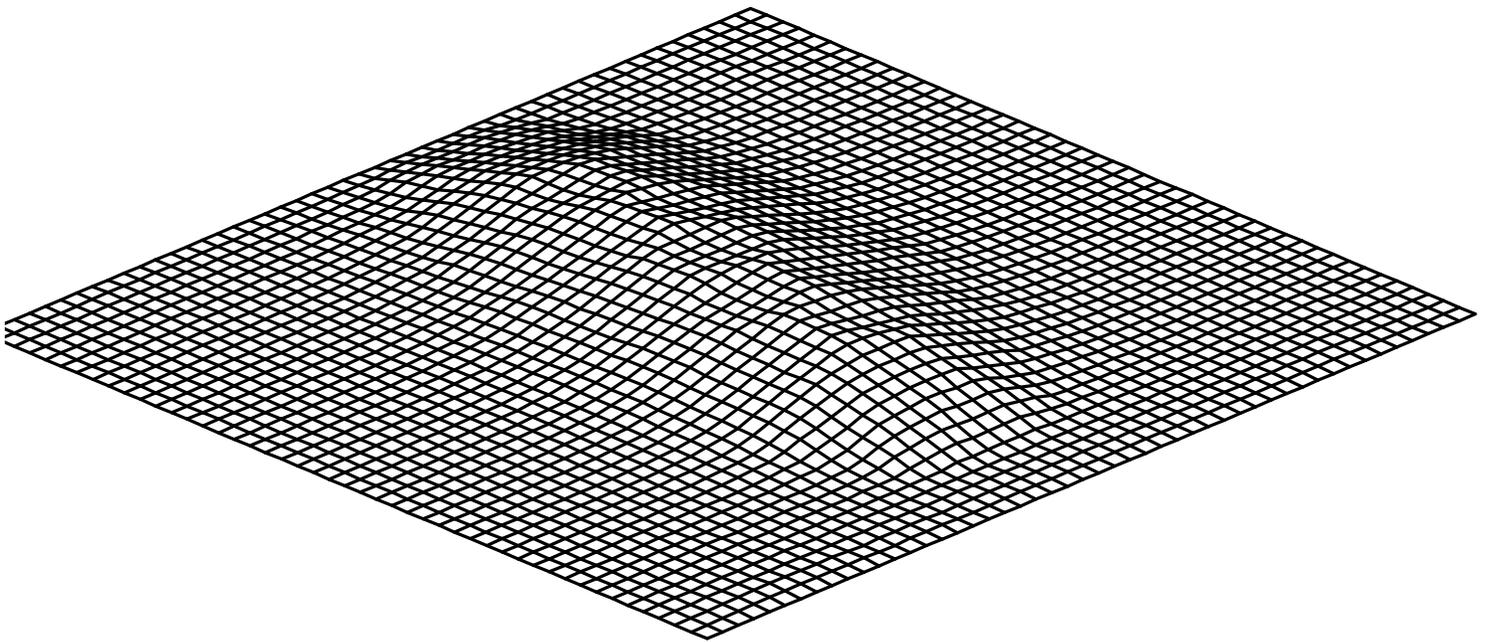
\end{document}